# Scaling to 32 GPUs on a Novel Composable System Architecture

John Ihnotic
Senior Director of Engineering
GigaIO
Carlsbad, California USA
jihnotic@gigaio.com

## ABSTRACT

The development of composable systems architecture[1] marks a significant shift in resource allocation and utilization within data centers. This paper presents a composable architecture scaling up to 32 GPUs on a single node, addressing the technical challenges encountered and the innovative solutions implemented. This design introduces a flexible and dynamic resource distribution mechanism, particularly for GPUs, enabling tailored allocation to meet varying node demands. The architecture's dynamic nature allows for the flexible assignment and reassignment of hardware resources, such as GPUs, to different nodes as required, offering unprecedented capability and flexibility.

## 1 Composable Systems Architecture

The technical challenge of scaling up a significant number of GPUs to a single node involves integrating 32 GPUs without modifying any existing code. This paper highlights the innovative approach taken to restructure traditional server resources into a more dynamic and user-centric model, enabling users to harness a substantial increase in computational power on a single node, tailored to their specific requirements. This composability not only facilitates ease of use but also expands the potential for computational tasks previously constrained by traditional servers. The composable systems architecture discussed is distinguished by its flexibility and capability to create configurations previously deemed impossible.

Figure 1. depicts a composable platform within a data center, contrasting traditional static infrastructure with a more dynamic, disaggregated approach. The left side shows a rack filled with fixed-function servers, whereas the rack on the right illustrates a flexible configuration with disaggregated components. Key advantages of this architecture include the ability to create highly configurable servers, such as a "32 GPU supercomputer," to deploy scalable solutions rapidly, to maximize resource utilization, and to enable heterogeneous computing.

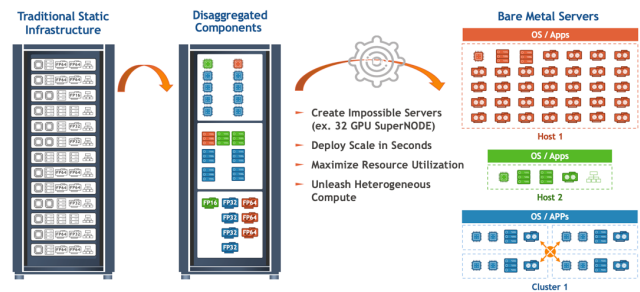

**Figure 1: Composable Platform in the Datacenter**

Figure 2 illustrates the memory fabric designed to carry all traffic in a data center environment. This fabric enables both server composition and inter-node communication, supporting various protocols like NVMe-OF, GDR, Libfabric, and MPI. It operates on a single PCIe network and promises no performance penalty, aiming to minimize the total cost of ownership (TCO) and improve serviceability. Additionally, it delivers scalability and easy integration, culminating in a rack-scale composition that is flexible enough to accommodate any server, or device, at any time.

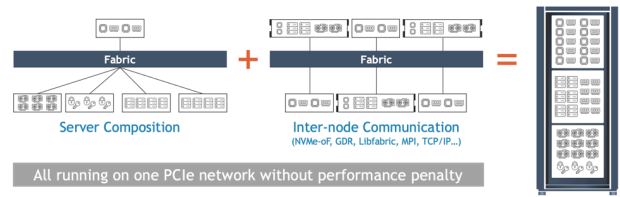

**Figure 2: One Universal AI Memory Fabric Carries All Traffic**

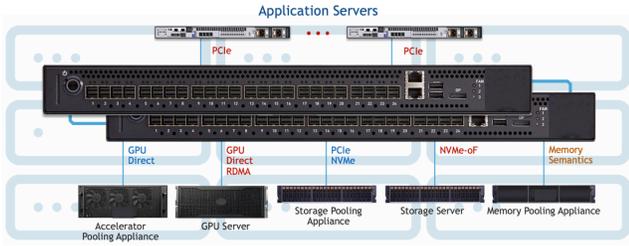

**Figure 3: Hardware and Connectivity**

Figure 3. shows the hardware and connectivity components of the composable infrastructure, focusing on the central role of the PCIe switch that interfaces with various elements like application servers, GPU servers, and storage solutions. The PCIe switch is designed to enable direct GPU communication (GPU Direct Memory Access), high-performance storage access (PCIe NVMe), and memory access (NVMe-oF and Memory Semantics).

A critical aspect of the system is the unified fabric which facilitates both traditional composability through a switching chip, and advanced inter-node communication channels, like those available with InfiniBand and high-speed Ethernet. This singular infrastructure enables the creation of a "server" connected to 32 GPUs, with the potential to extend this fabric substantially into a comprehensive data center solution.

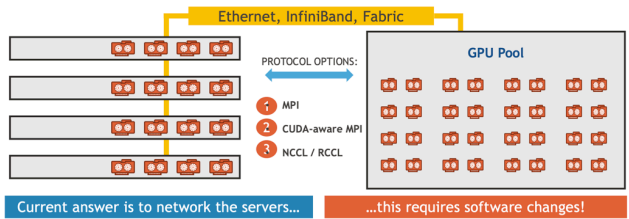

**Figure 4: AI + Accelerated Computing Before and After**

Figure 4. details the state of AI and accelerated computing before the introduction of this solution. The left side of the diagram illustrates a typical networking setup for 32 GPUs, arranged into eight servers, with four GPUs in each server, and interconnected through Ethernet or InfiniBand. Harnessing the power of these 32 GPUs currently requires communication using messaging protocols such as MPI (Message Passing Interface), CUDA-aware MPI, and/or NCCL/RCCL (NVIDIA Collective Communications Library/Radeon Collective Communications Library). These protocols are essential for communication between GPUs and are used to manage parallel processing tasks, but they add quite a bit of complexity, including software changes to distribute the code across servers. On the right side, all GPUs are aggregated onto a single fabric, on a single server, which presents technical challenges described in this paper. Figure 5. shows a functional layout of the novel 32 GPU server made possible from the composable architecture.

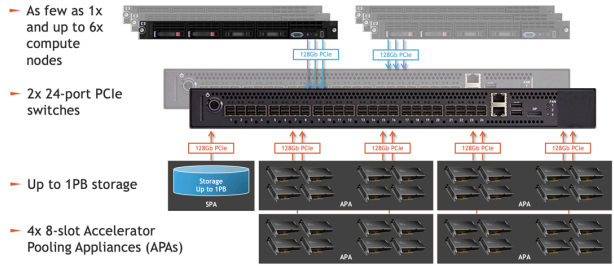

**Figure 5: 32 GPU Server Configuration**

## 2 Technical Challenges

Scaling to 32 GPUs presents several technical challenges, including BIOS enumeration, GPU driver support, and AI framework compatibility. Each challenge is dissected to understand the limitations of current system architectures and the innovative approaches taken to address them.

### 2.1 BIOS Enumeration

Addressing the BIOS enumeration challenge in GPU scaling involves accommodating a large memory footprint within modern system architectures. For example, integrating thirty-two 64GB Mi210 GPUs, which cumulatively require 2TB of memory, poses a substantial challenge due to the memory window limitations of CPUs. This challenge is exacerbated by vendor-specific BIOS configurations that may inefficiently allocate memory, potentially doubling the required system memory to accommodate GPU resources. Collaborative efforts with vendors were initiated to reconcile BIOS enumeration algorithms with the capabilities of contemporary CPU architectures, which are progressively supporting larger physical address spaces. This collaboration aims to ensure that system BIOS can support a multitude of devices, leveraging advancements in technology such as the advent of CXL. This development is anticipated to be a significant stride forward for the industry, allowing for more flexible and powerful composability solutions.

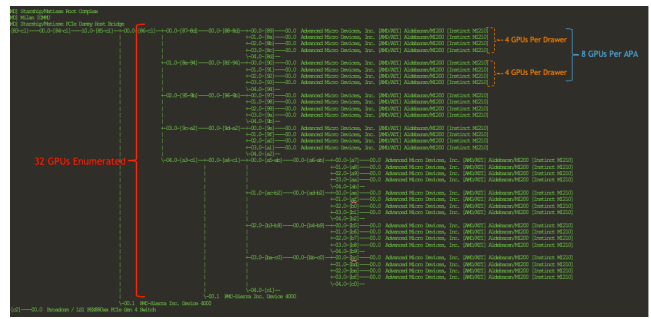

**Figure 6: 32 GPU LSPCI Tree**

Figure 6. shows an LSPCI tree output displaying the enumeration of 32 GPUs. On the left, is the primary PCIe root complex, and branching off of this are the GPUs organized into Advanced Micro

Devices, Inc. [AMD] nodes, specifically identifying each as an Alphacore/Mi200 [Instinct MI210]. The GPUs are further organized into groups, with annotations indicating "4 GPUs Per Drawer" and "8 GPUs Per APA," corresponding to the hierarchical physical arrangement within the system, and that all 32 GPUs have been successfully enumerated and are recognized by the system.

## 2.2 GPU Driver Support

GPU driver support is essential for utilizing the GPUs once recognized by the system. Driver support is also dependent on compatibility with various computing frameworks, like CUDA for NVIDIA or ROCm for AMD, which provide the runtime environments necessary for executing parallel computations on the GPUs. The drivers must be optimized to exploit the full bandwidth provided by the PCIe lanes, minimizing latency, and maximizing throughput for data-intensive tasks.

Engagement with vendors such as AMD and NVIDIA expanded the limits of GPU instances supported by their drivers (CUDA 12.3 and ROCm 5.7). The endeavor required strategic partnerships with those vendors to address and amend inherent driver restrictions, enabling support for an increased GPU count. Successful vendor collaboration led to driver updates being up-streamed, culminating in the latest driver releases supporting up to 64 GPUs—a milestone announcement. This breakthrough in driver support paves the way for future enhancements, projecting the possibility of even greater GPU scalability in subsequent system generations.

## 2.3 AI Framework Compatibility

AI frameworks like PyTorch and TensorFlow abstract away the complexity of directly interfacing with hardware, allowing developers to leverage GPUs for accelerated computing with minimal concern for the underlying infrastructure. This synergy requires the frameworks to be cognizant of the increased GPU capacity, a challenge met by advocating for changes within these ecosystems to accommodate the higher GPU counts found in advanced computing nodes. Through collaboration and open-source community engagement, adjustments were made to the frameworks' codebases (PyTorch 2.1 and TensorFlow 19.10), reflecting the new hardware capabilities, thus resolving issues, and expanding the operational limits for AI computations.

## 2.4 Containers

The adoption of containerization in AI development represents a significant leap forward in simplifying computational infrastructure. Containers encapsulate all the necessary components—code, runtime, system tools—into a single package, thereby abstracting the complexities of the underlying infrastructure. Utilizing repositories like NVIDIA GPU Cloud (NGC) or AMD Instinct's Docker container libraries, developers can deploy pre-built containers with optimized settings for high-performance AI applications. This approach streamlines workflows, allowing for rapid deployment and testing across various environments, and is crucial for demonstrating the capabilities of advanced computing systems without the overhead of traditional infrastructure setup.

## 3 Performance and Results

The performance of the composable systems architecture was explored through GPU-to-GPU peer-to-peer bandwidth tests. As shown in Figure 7., utilizing x8 connectivity, referred to as the "128 connection", the system facilitates efficient inter-GPU communication with minimal switching. Performance tests show P2P bandwidth reaching approximately 25 GB/s, against a theoretical maximum of 32 GB/s. This is attributed to a design that allows GPUs to bypass the CPU's root port complex, interfacing directly over a unified port, which significantly reduces latency and enhances data transfer rates.

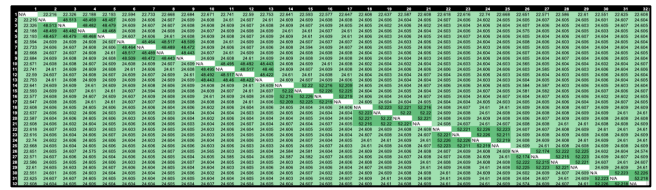

Figure 7: 32 GPU to GPU P2P Bandwidth

This peer-to-peer configuration substantially reduces the need for data to traverse the upper hierarchical layers of the CPU, thereby decreasing latency and enhancing overall performance. In conventional systems, GPU-to-GPU communication typically follows a multi-tiered path. The data must traverse from the source GPU through the node's CPU, then to the other CPU, and then move down through its PCIe switch to the target GPU. This hierarchical traversal, often involving multiple hops and the CPU's involvement, introduces latency and affects the overall efficiency of the GPU communication this system eliminates.

The theoretical performance explored above through bandwidth tests was then tested on two types of codes: an LLM for AI applications, and a CFD simulation for HPC workloads.

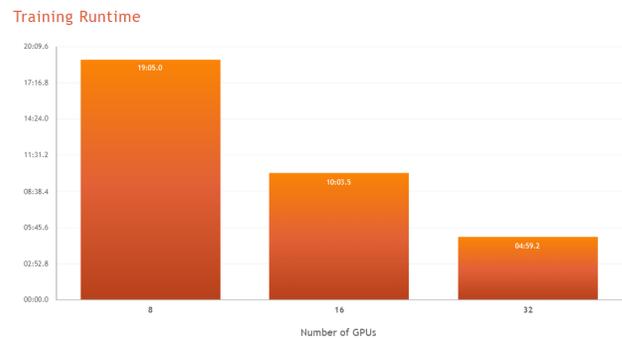

Figure 8: LLaMA 7B Training Runtime (Single Node)

Figure 8. illustrates the training runtime for a LLaMA[2] LLM model with 7 billion parameters across different GPU configurations on a single node. There are three bars, each representing the number of GPUs used: 8, 16, and 32. The first bar indicates that using 8 GPUs results in a training runtime of 19 hours and 5 minutes. The second bar shows a reduced runtime of 10 hours and 3.5 minutes when the number of GPUs is increased to 16. The third bar demonstrates a further reduction in training time to just 4 hours and 59.2 minutes with 32 GPUs demonstrating perfect scaling as the number of GPUs is increased.

For HPC applications, Figure 9. shows Dr. Moritz Lehmann's FluidX3D[3] CFD simulation of the Concorde's landing, resolving an immense 40 billion cell problem in just 33 hours using a composable system with 32 AMD Instinct MI210 accelerators. The factors making this possible include support for a high number of GPUs, substantial directly accessible GPU Memory VRAM of 2TB, and a single high-performance network fabric, which together drastically improve CFD simulation performance.

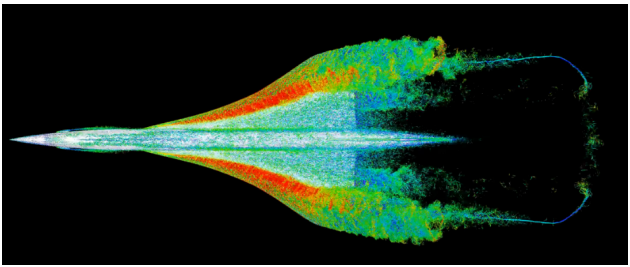

**Figure 9: Concorde Landing Simulation**

The immense GPU memory pool available with a 32-GPU system is specifically tailored for handling very large models, ideal for large datasets and complex AI models. It provides an extensive memory space for GPU computations, thereby minimizing the need for frequent data transfers between the GPU and system memory. This enables the processing of larger and more complex models with efficiency, significantly enhancing overall system performance and computational productivity.

## 4   Conclusion

The architecture described in this paper represents a significant leap in the realm of high-performance computing and AI, providing a robust and scalable solution for integrating a large number of GPUs into a single-node system. Through the innovative use of composable systems, this novel architecture has demonstrated the ability to scale to 32 GPUs without necessitating any changes to existing codebases, effectively overcoming significant technical barriers associated with BIOS enumeration, GPU driver support, and AI framework compatibility. This paper's findings have considerable implications for the future of AI and high-performance computing infrastructure. This architecture paves the way for enhanced computational capabilities within data centers, allowing for greater scalability and adaptability to the ever-increasing computational demands of modern AI and machine learning workloads.

## 5   Future Work

Future research directions include the integration of the Compute Express Link (CXL) into the composable system architecture. The roadmap for CXL integration includes the development of a new generation of hardware that can fully leverage this interconnect standard, offering greater bandwidth and reduced latency. particularly concerning memory pooling and coherence. CXL facilitates the aggregation of various memory types, offering much higher capacities than previously possible, and introduces memory coherency with CXL 3.0. This opens the door to rapidly composable infrastructure to meet the diverse needs of AI workloads. This advancement is expected to optimize the composability of resources further, enabling even more dynamic and efficient allocation of GPUs, memory, and other compute elements across nodes. Additionally, research will focus on creating robust management software for these enhancements, ensuring seamless operation and integration with existing data center infrastructures.